\documentstyle[amssymb,aps,prl,multicol,epsfig]{revtex}
\begin{document}
\title{Magnetism, Critical Fluctuations and Susceptibility Renormalization
in Pd}
\author{P. Larson, I.I.Mazin, and D.J. Singh}
\address{Center for Computational Materials Science,\\
Naval Research Laboratory, Washington, DC 20375-5000}
\date{\today }
\maketitle

\begin{abstract}
Some of the most popular ways to treat quantum critical materials, that is,
materials close to a magnetic instability, are based on the Landau
functional. The central quantity of such approaches is the average magnitude
of spin fluctuations, which is very difficult to measure experimentally or
compute directly from the first principles. We calculate the parameters of
the Landau functional for Pd and use these to connect the critical
fluctuations beyond the local-density
approximation and the band structure.
\end{abstract}

\begin{multicols}{2}
The physics and materials science of weak itinerant ferromagnetic metals and 
highly renormalized paramagnets near magnetic instabilities has attracted 
renewed theoretical interest. This is a result of recent discoveries of 
materials with highly non-conventional metallic properties, especially, 
non-Fermi liquid scalings, metamagnetic behavior, and unconventional 
superconductivity, in several cases co-existing with ferromagnetism. 
Discoveries in the last three years alone include the co-existing 
ferromagnetism and superconductivity of ZrZn$_{2}$\cite{pfleiderer}, 
UGe$_{2}$\cite{saxena},URhGe$_{2}$\cite{aoki}, high pressure $\epsilon 
$-Fe\cite{fe}, and the metamagnetic quantum critical point in 
Sr$_{3}$Ru$_{2}$O$_{7}$ \cite{grigera}.

Unfortunately, although model theories have been put forth, there is still
not an established material specific (first principles) theoretical
understanding of these phenomena. One difficulty is the usual starting point 
for first principles theories, density functional theory (DFT) as implemented 
in the local density approximation (LDA). This already includes most spin 
degrees of freedom, including dynamical fluctuations, as evidenced by its 
formally exact description of the uniform electron gas as well as its well 
documented success in accurately describing a wide variety of itinerant 
magnetic materials. However, the electron gas, upon which most density
functionals are built, is not near any critical point for densities relevant
to the solid state, and furthermore the proximity to itinerant magnetism of
a metal is an extremely non-local quantity, in particular depending on the
electronic density of states at the Fermi level $N(E_F)$.
Therefore, the exact DFT, which by definition includes all fluctuations and
describes the ground state magnetization exactly, is likely to be extremely
nonlocal and probably nonanalytical for the materilas near a quantum critical 
point.

On the other hand, the LDA, while providing a good description of most
itinerant ferromagnets that are not near critical points, fails to include
the soft critical fluctuations in the materials of interest here. Since
fluctuations are generically antagonistic to ordering, the result is that
magnetic moments and magnetic energies of weak itinerant ferromagnets near 
critical points are overestimated in the LDA, as opposed to LDA's failure to 
describe Mott-Hubbard insulators where the LDA {\it under}estimates the 
tendency to magnetism. Recent examples include Sc$_{3}$In\cite{sc3in}, 
Ni$_{3}$Al\cite{ni3ga-jap}, NaCo$_2$O$_4$\cite{singh00}, and 
ZrZn$_{2}$\cite{zrzn2}. Similarly, susceptibilities of paramagnets near 
critical points are underestimated. Furthermore, there is an overlap region 
where the LDA predicts ferromagnetism for paramagnetic materials. This 
interesting class includes FeAl\cite{singh,mm}, Ni$_{3}$Ga\cite{ni3ga-jap}, 
and Sr$_{3}$Ru$_{2}$O$_{7}$\cite{237} (as mentioned, this latter material 
shows a metamagnetic quantum critical point). The basic theoretical difficulty 
in correcting the LDA for these materials is that there is some unknown and 
possibly strongly material dependent cross-over in energy (and possibly 
non-trivially in momentum) separating quantum critical fluctuations, not 
included in the LDA, from the dynamical fluctuations that are included in the 
LDA. Qualitatively, this may be understood from the fact that the LDA is based 
on the properties of the uniform electron gas, which is far from any magnetic 
critical point at densities relavent for solids, the consequence being a 
mean-field-like description of magnetism near critical points. Thus the 
underlying reason for the failure of the LDA to describe these systems is very 
different from the failures in the well-known class of Coulomb correlated 
materials, such as the Mott-Hubbard insulators. There, the basic problem is 
the neglect of some electron-electron interactions, and can often be largely 
corrected at the static level, e.g. via approaches like LDA+U. It is worth 
noting that these dynamical fluctuations are responsible not only for the 
suppression of the magnetic ordering, but also for unusual transport 
properties of quantum critical materials, deviating from the conventional 
Fermi liquid behavior, for mass renormalization, and even for 
superconductivity in some systems. Many of these issues have been addressed 
recently in theoretical papers, utilizing idealized models of various kinds. 
However, a quantitative link between such models and actual material 
characteristics is still missing.

We attempt to build a bridge between such theories and the LDA. We concentrate 
on the question of what kind of material-specific understanding, relevant for 
quantum criticality, can be extracted from the LDA calculations. Primarily, 
we focus here on Pd. This is perhaps the best studied high susceptibility 
paramagnet\cite{FA,JF,OZD,SA}, and in fact a number of theories related to 
spin fluctuations have been elucidated using this material. Furthermore, 
itinerant ferromagnetism appears in Pd at 2.5\% Ni doping\cite{PdNi}. We 
present highly accurate calculations of the static magnetic susceptibility for 
Pd and find that, indeed, the LDA overestimates the tendency to magnetism. We 
also estimate the r.m.s. magnitude of spin fluctuations (paramagnons) in Pd, 
needed to reduce the calculated susceptibility to reproduce experiment, and 
show that it is compatible with that which might be estimated from LDA 
susceptibility $via$ the fluctuation-dissipation theorem with a reasonable 
ansatz for the cut-off momentum.

We have performed electronic structure calculations using the self
consistent full potential linearized augmented plane wave (FLAPW)\cite%
{singh2} method within the density functional theory (DFT)\cite{kohn}. The
local density approximation (LDA) of Perdew and Wang\cite{perdewwang} and
the Generalized Gradient Approximation (GGA) of Perdew, Burke, and Ernzerhof%
\cite{perdew} were used for the correlation and exchange potentials.
Calculations were performed using the WIEN2k package\cite{wien2k}. Local
orbital extensions\cite{LAPW1} were included in order to accurately treat
the upper core states and to relax any residual linearization errors. A well
converged basis consisting of LAPW basis functions with wave vectors up to $%
K_{\max }$ set as $RK_{\max }=9,$ with the Pd sphere radii $R=$2.59 bohr. All 
total energy calculations used at least 1470 and up to 2844 {\bf k}-points in 
the irreducible part of the Brillouin zone as needed. Spin-orbit (SO) 
interactions were incorporated using a second variational 
procedure\cite{spinorb}, where all states below the cutoff energy 1.5 Ry were 
included, with the so-called $p_{1/2}$ extension\cite{singh2}, which accounts 
for the finite character of the wave function at the nucleus for the $p_{1/2}$ 
state.

All calculations were performed in an external magnetic field, interacting
with both spin, {\bf s,} and orbital, {\bf l,}\cite{wien2k} momenta: 
\[
V_{H_{ext}}=\mu _{B}{\bf H}_{ext}{\bf ^{.}}({\bf l}+2{\bf s}). 
\]%
The input values of $H$ were chosen from 0 to 10000 T in irregular
increments to map out the change in energy and magnetic moment as a function
of applied field. While use of the LDA\cite{perdewwang} resulted in zero
magnetic moment in a zero magnetic field, consistent with the experiment\cite%
{PdNi}, use of GGA\cite{perdew} resulted in a persistent magnetic moment of 
0.2$\mu_{B}$, with an extremely small magnetic energy of less than 1 meV. 

\begin{figure}[htb]

\centerline{
\epsfig{file={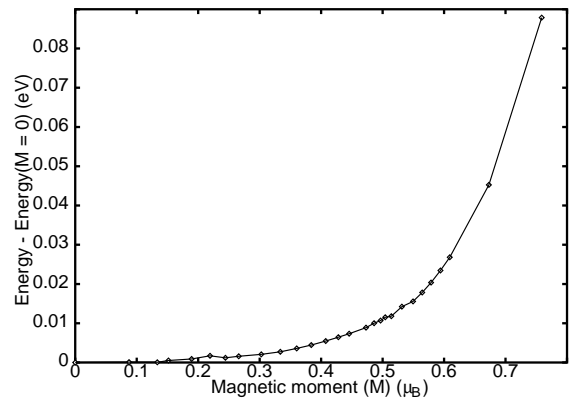}, width= 0.9\linewidth}}
\caption{Calculated LSDA total energy, E, (in eV) with respect to M = 0$\protect\mu_{B}$ as a function of calculated magnetic moment, M (in $\protect\mu_{B}$).}
\label{energy}
\end{figure}

\begin{figure}[htb]

\centerline{
\epsfig{file={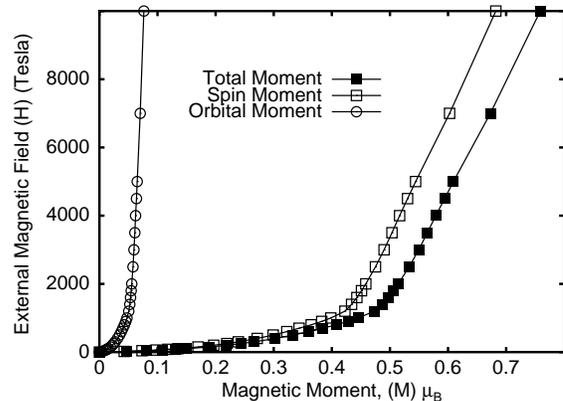}, width= .9\linewidth}
}
\caption{Applied external magnetic field, H, (in Tesla) as a function of the
calculated LSDA magnetic moment, M (in $\protect\mu_{B}$). The total moment is 
shown together with spin component and the orbital component. }
\label{mh}
\end{figure}

In order to understand the change in the total energy and magnetic moments
as a function of the applied external field, special care was taken to
ensure that these quantities were well converged with respect to the 
{\bf k}-mesh. Given that in the low fields we are interested in, energy 
changes need to be converged of the order of 0.1 meV/atom. The total energy, 
$E$, with respect to that at $M=0$ $\mu _{B}$ as a function of the 
magnetization, $M$, is shown in Figure \ref{energy}. Figure \ref{mh} shows 
the applied magnetic field, $H$, as a function of $M$ (with the magnetization 
direction 100). Note that the latter dependence follows from the former one, 
as $H\equiv \frac{\partial E}{\partial M}$. One can see though that of the two 
quantities $H$ shows less computational noise, so this was the dependency we 
used in the analysis described below.

As can be seen in both plots (more so in Figure \ref{mh}), there exist two
regimes in terms of the magnetic moment, $M$. For values of $M$ $\leq 0.5$ $%
\mu _{B}$ (corresponding to $H$ $\sim $ 1200 T), the external field and
energy increase slowly, but for M $\geq $ 0.5$\mu _{B}$, both $H$ and $E$
increase rapidly, suggesting that the long wave spin fluctuations at any
temperature should be smaller that $\sim 0.5$ $\mu _{B}$ in amplitude.

\begin{figure}[htb]
\centerline{
\epsfig{file={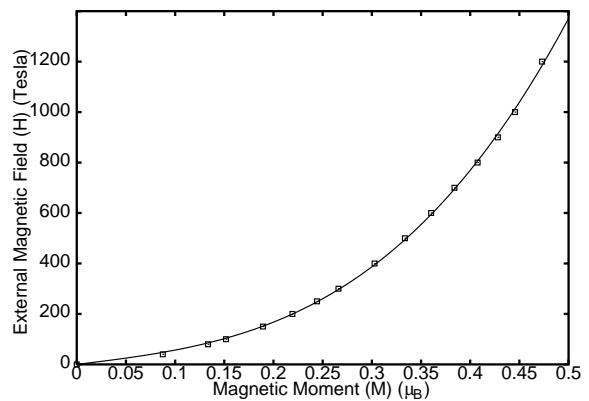}, width= .9\linewidth}
}
\caption{The external magnetic field, H, (in Tesla) as a function of the
calculated magnetic moments, M (in $\protect\mu_{B}$). The fit is to n $\leq$
3 in Eqn. \protect\ref{Hexp}
}
\label{fit}
\end{figure}

\begin{figure}[htb]
\centerline{
\epsfig{file={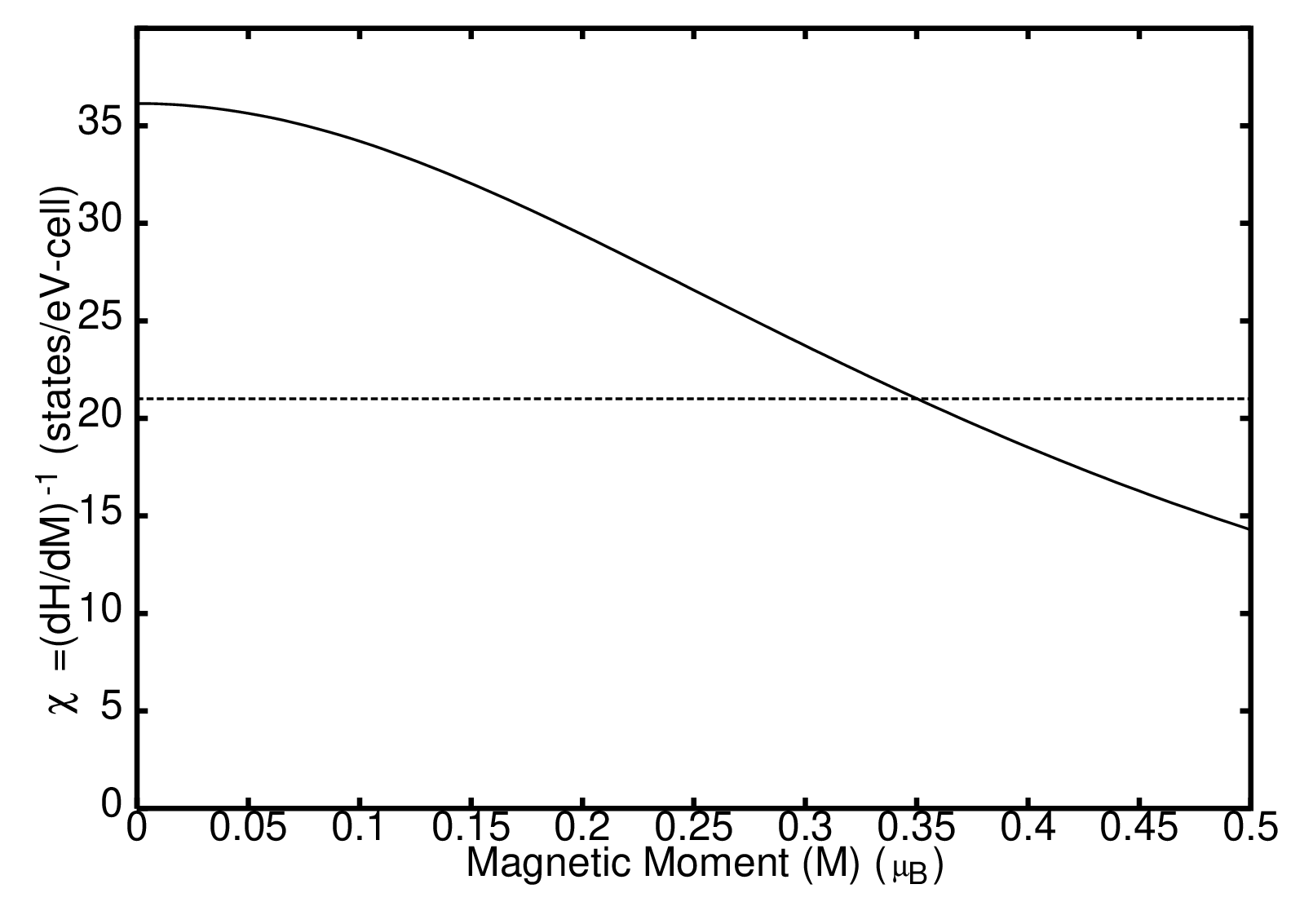}, width= .9\linewidth}
}
\caption{Magnetic susceptibility, $\protect\chi$, (in states/eV-cell)
calculated from the fit of H, $\chi$ = $(\frac{\partial H}{\partial M})^{-1}$, 
shown as a function of M. The dashed line at 21 states/eV-cell corresponds
to the experimental value of $\protect\chi$ for Pd\protect\cite{Foner,Pd2}.}
\label{movh}
\end{figure}

The linear magnetic susceptibility is defined as $\chi ^{-1}=
\frac{\partial H}{\partial M}|_{M=0}=\frac{\partial ^{2}E}{\partial M^{2}}$. 
Figure \ref{fit} shows, however, that even for $M\lesssim 0.5$ $\mu _{B}$ the 
susceptibility is highly nonlinear. In fact, $\frac{\partial M}{\partial H}$ 
starts near 11.6$\times 10^{-4}$ emu/mol and decreases rapidly with the field. In order to 
compute accurately the relevant derivatives, we have fitted the calculated 
$H(M)$ for $M<0.5$ $\mu _{B}$ with a polynomial (Figure \ref{fit}). Thus 
computed susceptibility as the function of the applied field is shown in Fig. 
\ref{movh}. We see that the zero field susceptibility is nearly twice larger 
than the experimental value of 6.8 $\times 10^{-4}$ emu/mol corresponding to 21 
st/eV-cell\cite{Foner,Pd2}. Only in a field of 550 T does the susceptibilty 
eventually become close to the experimental number.

One may understand the origin of this overestimation of
magnetic susceptibility in the following way. Not only is the calculated
susceptibility very large, but also as mentioned the dependence of the induced 
magnetic moment on the applied field is highly nonlinear in such a manner that 
the total energy as a function of the constrained magnetic moment is very flat
up to $M\approx 0.5$ $\mu _{B}.$ This implies that zero temperature
quantum fluctuations beyond the LDA
may have a substantial magnitude. One of the ways to
take into account these fluctuations is $via$ the Ginzburg-Landau theory,
which, in connection with the spin fluctuations in nearly-magnetic metals
has been used by several authors during the 1970's. This method
starts with an expression for the total energy {\it without} such
fluctuations as a function of the induced magnetic moment $M$ 
\begin{eqnarray}
E_{static}(M) &=&a_{0}+\sum_{n\geq 1}\frac{1}{2n}a_{2n}M^{2n},  \label{Eexp}
\\
H_{static}(M) &=&\sum_{n\geq 1}a_{2n}M^{2n-1}  \label{Hexp}
\end{eqnarray}%
(obviously, $a_{2}$ gives the inverse spin susceptibility without 
fluctuations), and then assume Gaussian zero-point fluctuations of an r.m.s. 
magnitude $\xi $ for each of the $d$ components of the magnetic moment (for
a 3D isotropic material like Pd, $d=3).$ After averaging over the spin 
fluctuations, one obtains a fluctuation-corrected functional. The general 
expression of Ref. \cite{Shimizu81} can be written in the following compact 
form:
\begin{eqnarray}
H(M) &=&\sum_{n\geq 1}\tilde{a}_{2n}M^{2n-1}  \nonumber \\
\tilde{a}_{2n} &=&\sum_{i\geq 0}C_{n+i-1}^{n-1}a_{2(n+i)}\xi ^{2i}\Pi
_{k=n}^{n+i-1}(1+\frac{2k}{d}).  \label{renorm}
\end{eqnarray}%
For instance,%
\begin{eqnarray}
\tilde{a}_{2} &=&a_{2}+\frac{5}{3}a_{4}\xi ^{2}+\frac{35}{9}a_{6}\xi ^{4}+%
\frac{35}{3}a_{8}\xi ^{6}...  \nonumber \\
\tilde{a}_{4} &=&a_{4}+\frac{14}{3}a_{6}\xi ^{2}+21a_{8}\xi ^{6}... 
\nonumber \\
&&...
\end{eqnarray}

We can now make a connection between the above theory and the band structure.
Our calculations, fitted to Eq. \ref{Hexp} with $n=3$, are presented in Fig. 
\ref{fit}. Since the high-power coefficients are positive, obviously,
renormalization according to Eq. \ref{renorm} will lead to a reduction of
the magnetic susceptibility, $\chi =1/\tilde{a}_{2}<1/a_{2}$. The magnitude of
this effect depends on the r.m.s. amplitude, $\xi ,$ of the spin fluctuations, 
which in turn depends on how fast  $\chi(q)$ changes at small $q$'s.

In order to find the value of $\xi $ necessary to renormalize the zero-field
value of $\chi $, one can use Eq. \ref{Hexp} with the n $\leq $ 3 expansion:%
\begin{equation}
\chi ^{-1}(0)=\frac{\partial M}{\partial H}=\tilde{a}_{2}=a_{2}+\frac{5}{3}a_{4}\xi^{2}+\frac{35}{9}a_{6}\xi ^{4}.
\end{equation}%
The fit coefficients are $a_{2}$ = 478 $T/\mu_{B}$, $a_{4}$ = 8990 
$T/\mu_{B}^{3}$, and $a_{6}$ = 277 $T/\mu_{B}^{5}$. Setting $\chi $(0) equal 
to the experimental value\cite{Foner,Pd2} leads to $\xi $ = 0.15$\mu _{B}$. 
However,it is highly desirable to find a way of estimating $\xi$ in a real 
material using $ab$ $initio$ calculations. This can be done using the 
fluctuation-dissipation theorem along the lines suggested by 
Moriya\cite{moriya} and elaborated by many authors (see, e.g., Refs. 
\cite{sol,kaul,IM}), which states that for zero-point fluctuations 
\begin{equation}
\xi ^{2}=\frac{4\hbar }{\Omega }\int d^{3}q\int \frac{d\omega }{2\pi }\frac{1%
}{2}%
\mathop{\rm Im}%
\chi ({\bf q},\omega ), \label{mumu}
\end{equation}%
where $\Omega$ is the Brillouin zone volume\cite{chi}.
 It is customary to approximate 
$\chi ({\bf q},\omega )$ near a QCP as%
\begin{equation}
\chi ^{-1}({\bf q},\omega )=\chi _{0}^{-1}(0,0)-I+cq^{2}-i\omega /\Gamma q,
\label{expchi}
\end{equation}%
where $\chi _{0}^{-1}(0,0)=1/N(E_{F})$ (density of states per spin) is the
bare (noninteracting) static uniform susceptibility, and $I$ is the Stoner
parameter which is weakly dependent on {\bf q} and $\omega$. Obviously, 
$\chi_{0}^{-1}({\bf q},\omega )=\chi _{0}^{-1}(0,0)+cq^{2}-i\omega /\Gamma q$ 
is the noninteracting susceptibility. Although not necessary\cite{kaul}, a 
convenient approximation, good near a QCP, is that 
$\chi^{-1}(0,0)\approx 0$, that is, $I\approx 1/N(E_F)$. One can also use an 
expansion for $\chi _{0}({\bf q},\omega ),$ equivalent to Eq. \ref{expchi}, 
namely%
\begin{equation}
\chi _{0}({\bf q},\omega )=N(E_{F})-aq^{2}+ib\omega /q.\label{8}
\end{equation}

Moriya mentioned in his book\cite{moriya} that the 
coefficients $a$ and $b$ are related, in some 
approximation, to the band structure, in particular, to the
effective mass of electrons at the Fermi level and to some contour integral
along a line on the Fermi surface. While Moria's expressions are
difficult to evaluate numerically within the standard band structure
calculations, one can rewrite equivalent
 expressions, better suited for actual calculations. For completness,
we present below the full derivation:
\end{multicols}
\begin{eqnarray}
\mathop{\rm Re}%
\chi _{0}({\bf q,}0) &=&\sum_{{\bf k}}\left[ f(E_{{\bf k}})-f(E_{{\bf k+q}})%
\right] (E_{{\bf k+q}}-E_{{\bf k}})^{-1}  \label{Re} \\
\mathop{\rm Im}%
\chi _{0}({\bf q,}\omega ) &=&\sum_{{\bf k}}[f(E_{{\bf k}})-f(E_{{\bf k+q}%
})]\delta (E_{{\bf k+q}}-E_{{\bf k}}-\omega ),\label{Im}
\end{eqnarray}%
where $f(E)$ is the Fermi function, $-\frac{df(E)}{dE}=\delta (E-E_{F})$. 
Expanding Eq. \ref{Re} in 
$\Delta =E_{{\bf k+q}}-E_{{\bf k}}={\bf v}_{{\bf k}}{\bf \cdot q+}\frac{1}{2}\sum_{\alpha \beta }\mu _{{\bf k}}^{\alpha \beta }q_{\alpha
}q_{\beta }+...,$ we get to second order in q%
\begin{equation}
\mathop{\rm Re}%
\chi _{0}({\bf q,0})=N(E_{F})+\sum_{{\bf k}}\left[ \frac 12\left( \frac{d\delta
(\varepsilon _{{\bf k}}-E_{F})}{dE_{F}}\right) ({\bf v}_{{\bf k}}{\bf \cdot
q+}\frac{\sum_{\alpha,\beta} \mu _{{\bf k}}^{\alpha \beta }q_{\alpha }q_{\beta }}{2}%
)+\frac 16 \left( \frac{d^{2}\delta (\varepsilon _{{\bf k}}-E_{F})}{dE_{F}^{2}}%
\right) ({\bf v}_{{\bf k}}{\bf \cdot q})^{2}\right] .
\end{equation}%
The odd powers of ${\bf v_k}$ cancel out and we get ($\alpha, \beta = x, y, z$)%
\begin{eqnarray}
%TCIMACRO{\func{Re}}%
%BeginExpansion
\mathop{\rm Re}%
%EndExpansion
\chi _{0}({\bf q}) &=&N(E_{F})+\sum_{\alpha,\beta}\frac{q_{\alpha }q_{\beta }%
}{4}\frac{d\left\langle N(E_{F})\mu ^{\alpha \beta }\right\rangle }{dE_{F}}+
\sum_{\alpha,\beta}\frac{q_{\alpha }q_{\beta }}{6}\frac{d^{2}\left\langle %
N(E_{F})v_{\alpha }v_{\beta}\right\rangle }{dE_{F}^{2}} \\
&=&N(E_{F})
+\frac{q^{2}}{4}\frac{%
d\left\langle N(E_{F})\mu _{xx}\right\rangle }{dE_{F}}
+\frac{q^{2}}{6}\frac{d^{2}\left\langle
N(E_{F})v_{x}^{2}\right\rangle }{dE_{F}^{2}}
 ,
\end{eqnarray}%
\begin{multicols}{2}
where $v_{x}^{2}=v_{y}^{2}=v_{z}^{2},$ $\mu _{xx}=\mu _{yy}=\mu _{zz}.$ The
last equality assumes cubic symmetry; generalization to a lower symmetry is
trivial. Using the following relation,%
$$
\sum_{{\bf k}}{\bf \nabla }_{{\bf k}}F(\varepsilon _{{\bf k}})=\sum_{{\bf k}}%
\frac{dF(\varepsilon _{{\bf k}})}{d\varepsilon _{{\bf k}}}{\bf \nabla }_{%
{\bf k}}\cdot \varepsilon _{{\bf k}}=\sum_{{\bf k}}\frac{dF(\varepsilon _{%
{\bf k}})}{d\varepsilon _{{\bf k}}}{\bf v}_{{\bf k}},
$$
one can prove that%
\begin{equation}
\frac{d^{2}\left\langle N(E_{F})v_{x}^{2}\right\rangle }{dE_{F}^{2}}=-\frac{%
d\left\langle N(E_{F})\mu _{xx}\right\rangle }{dE_{F}}.
\end{equation}%
Therefore 
\begin{equation}
\mathop{\rm Re}%
\chi _{0}({\bf q})=N(E_{F})-\frac{q^{2}}{12}\frac{d^{2}\left\langle
N(E_{F})v_{x}^{2}\right\rangle }{dE_{F}^{2}}  \label{rechi}
\end{equation}

Similarly, for Eq. \ref{Im} one has 
\begin{equation}
\mathop{\rm Im}%
%EndExpansion
\chi _{0}({\bf q,}\omega )=\sum_{{\bf k}}\left[ \left( -\frac{df(\varepsilon
)}{d\varepsilon }\right) \omega \delta ({\bf v}_{{\bf k}}{\bf \cdot q}%
-\omega )\right]
\end{equation}

After averaging over the directions of ${\bf q,}$ this becomes, for small $%
\omega ,$%
\begin{eqnarray}
\mathop{\rm Im}%
%EndExpansion
\chi _{0}(q{\bf ,}\omega ) &=&\frac{\omega }{2}\sum_{{\bf k}}\frac{\delta
(\varepsilon _{{\bf k}})}{v_{{\bf k}}q}\theta (v_{{\bf k}}q-\omega )=\frac{%
\omega }{2q}\left\langle N(E_{F})v^{-1}\right\rangle \nonumber \\
v &=&\sqrt{v_{x}^{2}+v_{y}^{2}+v_{z}^{2}}.
 \label{imchi} 
\end{eqnarray}%
Although the Fermi velocity is obviously different along different directions,
it is still a reasonable approximation to introduce an average $v_{F}$.
Then the frequency cutoff in Eq. \ref{imchi} is $\omega _{c}\approx qv_{F}$.

From Eq. \ref{8} it follows that
\begin{equation}
\mathop{\rm Im}%
%EndExpansion
\chi ({\bf q,}\omega )=\frac{bq\omega N(E_{F})^{2}}{a^{2}q^{6}+b^{2}\omega ^{2}},
\end{equation}%
and, performing the integrations,%
\begin{eqnarray*}
\xi ^{2} &=&\frac{bv_{F}^{2}N(E_{F})^{2}}{2a^{2}\Omega }[Q^{4}\ln
(1+Q^{-4})+\ln (1+Q^{4})] \\
&=&\frac{3b\left\langle N(E_{F})v_{x}^{2}\right\rangle N(E_{F})}{%
2a^{2}\Omega }[Q^{4}\ln (1+Q^{-4})+\ln (1+Q^{4})],
\end{eqnarray*}%
where $Q=q_{c}\sqrt{\frac{a}{bv_{F}}}$ with $q_{c}$ the cutoff in the 
momentum space. There is no solid prescription to estimate the cutoff 
value. At small $Q$ the dependence of $\xi$ on $Q$ is quadratic,
however, at large $Q$ it becomes relatively weak (logarithmic).
While the susceptibility $\chi({\bf q},\omega)$ can, in principle, be 
calculated exactly, there is no rigorous definition of $q_c$. The conceptual 
difficulty here is, as in all problems related to electron-electron 
interactions, that some part of the effect in question is already included in 
the LDA, and rigorous treatment of the double-counting becomes virtually 
impossible (cf. discussion of this issue in connection to the LDA+U 
method\cite{LDAU}). At this point one needs to make some choice of $q_{c}$. 
A natural ansatz is to 
choose the value of $q$ at which the model susceptibility (Eq. 
\ref{rechi}) becomes unphysical (negative), 
$q_{c}=\sqrt{\frac{N(E_{F})}{a}}$. 

The above formulas reduce all parameters needed for estimating the 
r.m.s. amplitude of spin fluctuations to four integrals over the Fermi 
surface: $N(E_{F})$, 
$a=\frac{1}{12}\frac{d^{2}\langle N(E_{F}) v_{x}^{2}\rangle}{dE_{F}^{2}}$
$b=\frac{1}{2}\left\langle N(E_{F})v^{-1}\right\rangle$,
$v_{F}=\sqrt{3\frac{\left\langle N(E_{F})v_{x}^{2}\right\rangle}{N(E_{F})}}$.  
It should be noted that these integrals are extremely sensitive to 
the {\bf k}-point mesh. We used various meshes between 40x40x40 and 60x60x60, 
and averaged the results using the bootstrap method\cite{boot}
 (to eliminate the effect of
special points coinciding with mesh points). Velocities were 
calculated\cite{velnote} as
matrix elements of the momentum operator, using the {\it optic }program of
the WIEN package. We obtained (all energies are measured in Ry, lengths in
Bohr, and velocities in Ry$\cdot $Bohr) $N(E_{F})$ = 17.1, 
$\left\langle N(E_{F})v_{x}^{2}\right\rangle$ = 0.58, 
$\frac{d^{2}\langle N(E_{F})v_{x}^{2}\rangle}{dE_{F}^{2}}$ = 1700,
$\left\langle N(E_{F})v^{-1}\right\rangle$ = 135,
$v_{F}=\sqrt{3\frac{\left\langle N(E_{F})v_{x}^{2}\right\rangle}{N(E_{F})}}$ = 
0.31. Correspondingly, 
$a\approx$ 140, $b\approx$ 72, and 
$q_{c}=\sqrt{\frac{N(E_{F})}{a}}=0.35$, using the above-mentioned
ansatz.

Now we get
\begin{equation}
\xi 
=0.2\mu _{B}\sqrt{Q^{4}\ln (1+Q^{-4})+\ln (1+Q^{4})},
\end{equation}%
and with
$Q=0.88$, we obtain  $\xi =$0.16 $\mu _{B}.$ Note 
that the energy of a long-range spin fluctuation with such an amplitude is of 
the order of a few meV per atom, as can be seen from Fig. \ref{energy}. 

\begin{figure}[htb]
\centerline{
\epsfig{file={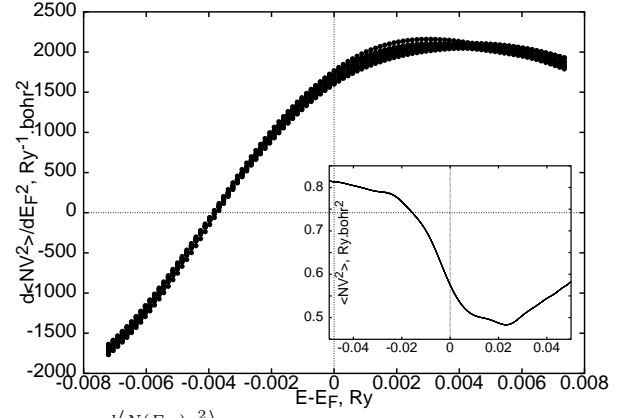}, width= .9\linewidth}
}
\caption{$\frac{d\left\langle N(E_{F})v_{x}^{2}\right\rangle}{dE_F}$ as a 
function of energy, calculated by the bootstrap method. Note the numerical 
noise of up to 10\%. Inset: $\left\langle N(E_{F})v_{x}^{2}\right\rangle $.}
\label{Nv2}

\end{figure}

This result is quite sensitive to the second derivative 
$\frac{d^{2}\left\langle N(E_{F})v_{x}^{2}\right\rangle}{dE_{F}^{2}}$, which 
was the most difficult quantity to calculate. An inspection of the energy 
dependence of $\left\langle N(E_{F})v_{x}^{2}\right\rangle $ (Fig. \ref{Nv2}, 
inset) elucidates the reason: the Fermi energy in Pd lies near an inflection 
point. As a result, 
$\frac{\ d^{2}\left\langle N(E_{F})v_{x}^{2}\right\rangle}{dE_{F}^{2}}$ is
small (and hard to calculate reliably). This, perhaps, is not accidental;
were this derivative 2-3 times larger, the mean amplitude of spin
fluctuation would have been relatively small even given extreme proximity of
this material to the ferromagnetic instability, because the relevant phase
space would have been too small. If this approximation is correct, this gives 
an important hint for identifying quantum critical materials from the LDA 
calculations: the calculated ground state should be close to ferromagnetic 
instability (on either side) {\it and} the Fermi energy should be close to an 
inflection point of the $\left\langle N(E)v_{x}^{2}\right\rangle $. 

The calculated value of $\xi $, if substituted into Eq. \ref{renorm}, gives 
$\chi \approx 6.4\times 10^{-4}$ emu/mol,
practically the same as the experimenatal number.
Such a good agreement is without doubt fortuitous; for instance, using
GGA as a starting point instead of LDA would have destroyed this agreement.\cite{GGA} 
We should keep in mind that, first of all, the formalism itself 
is very crude;  $\chi _{0}(q,\omega )$ was expanded to leading terms at small 
$q$, but this expansion is used up to some large $q_c$ comparable with $k_F$. 
Furthermore, a key parameter in the formalism is the cut-off momentum $q_c$, 
for which we use an ansatz based on the {\it large}-$q$ behavior of the model 
$\chi (q,\omega )$.

However, the fact that this procedure produces a correction of the right
order of magnitude is probably robust and suggests that the underlying
physics was identified correctly.

{To summarize, we use highly accurate LDA calculations
to estimate the parameters in Moriya's
spin fluctuation theory, and thereby estimate the corrections,
due to long wavelength spin fluctuations, to the LDA results.
Let us, in conclusion,} repeat our main points. The key parameter defining the 
nontrivial physics near the QCP is the mean-square amplitude of the spin 
fluctuations. This parameter is a highly material dependent, nonlocal 
quantity, determined by the spin susceptibility in a large part of the 
Brillouin zone, as well as by the characteristic cut-off length separating
``non-trivial'' spin fluctuations from spin-fluctuation implicitly included in
the LDA. It is hoped, however, that this parameter is mainly defined by 
the long wavelength part of the susceptibility, while the short wave-length 
characteristics, including the cut-off length, may be only weakly material, 
pressure, $etc$., dependent. We implements this idea, 
 relating, in the corresponding approximation, the mean-square amplitude of
the spin fluctuations near a QCP with characteristics of the one-electron
band structure. The formalism is based on the (1) Stoner theory for spin
susceptibility, (2) fluctuation-dissipation theorem, and (3) lowest-order
expansion of the real and imaginary part of the polarization operator in
terms of the frequency and the wave vector. {The actual band structure
of the material is taken into account $via$ the lowest-order expansion
coefficients of the LDA susceptibility, while the effects beyond the
lowest order in $q$ and $\omega$ are neglected.}  
Together with the Landau
expansion of the free energy, also computable within the LDA formalism, this
allows one to treat quantum criticality semi-quantitatively on the basis of
LDA calculations. 

We are grateful for helpful discussions with A. Aguayo, A. Chubukov, S. 
Halilov, G. Lonzarich, and S. Saxena. Work at the Naval Research Laboratory is 
supported by the Office of Naval Research.

\begin{figure}[htb]
\centerline{
\epsfig{file={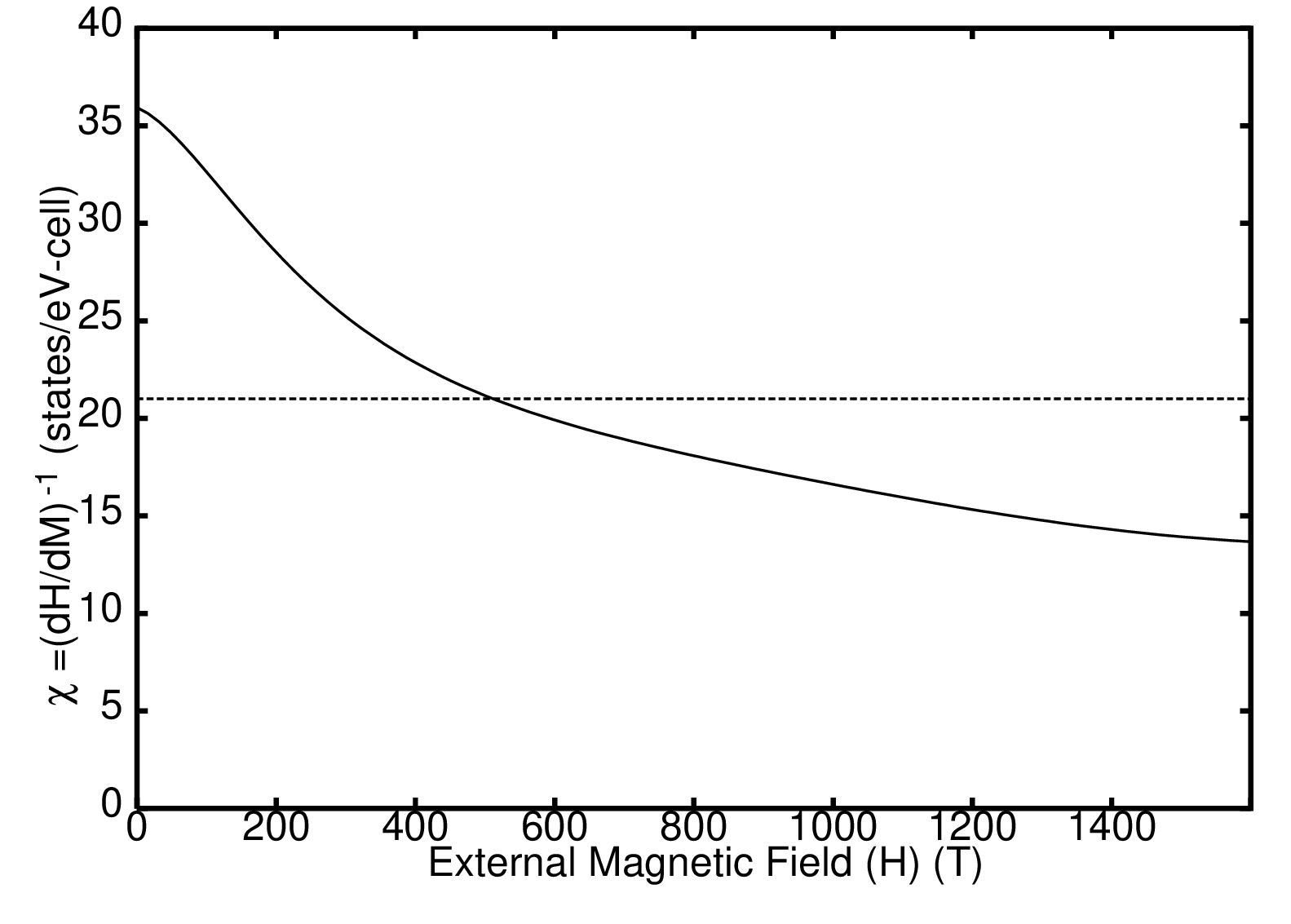}, width= .9\linewidth}
}
\caption{Magnetic susceptibility, $\protect\chi$, (in states/eV-cell)
calculated from the fit of H, $\chi$ = $(\frac{\partial H}{\partial M})^{-1}$,
shown as a function of H. The dashed line at 21 states/eV-cell corresponds
to the experimental value of $\protect\chi$ for Pd\protect\cite{Foner,Pd2}.}
\label{movh1}
\end{figure}
\end{multicols}

\begin{references}
\bibitem{chi} {Despite close similarity, what we propose is not
exactly the same as the formalism developed in Refs.\cite{moriya,sol,%
kaul} and others. The latter starts from a Landau functional which
does not include any fluctuations at all, while we start from  LDA
 energies which includes some, but not all fluctuations (if we had
used the exact Hohenberg-Kohn theory we would not need any
correction at all). Therefore $\chi $ entering
Eq. \ref{mumu} formally is not the experimental susceptibility 
as in the above-mentioned works.}
\bibitem{GGA} We have not performed GGA calculations at the same level of 
accuracy as LDA ones; however, GGA fixed spin moment calculations without the 
spin-orbit interaction indicate that correcting the spin susceptibility to the
experimental value would require $\xi\approx 0.27$ $\mu_{B}$,  larger but
comparable with 
the value of 0.16 $\mu_{B}$ calculated above. Obviously, while neither 
LDA nor GGA cannot properly account for the long-range critical fluctuations, 
the question which approximation describes the non-critical physics better
is open. In fact, the same question often arises in more standard band
structure calculations as well.
\bibitem{velnote} Fermi velocity in Pd is renormalized by electron-paramagnon
interaction in much the same way as by electron-phonon interaction. It is not 
entirely clear whether this renormalization should be taken into account when 
applying the fluctuation-dissipation theorem. Fortunately, this 
renormalization is not large enough to change the results in any important 
way; indeed, standard LDA calculations for optical conductivity, including 
plasma frequency, agree well with the experiment (see, e.g., Ref. \cite{opt}). 
Calculated DC transport agrees with the experiment within 15\% \cite{SS}. The 
linear specific heat coefficient, usually the most affected quantity, also 
deviates from the calculated value of 8.2 mJ/mol K$^2$ (using our $(N(E_F)$ 
and electron-phonon coupling constant of 0.35 from Ref. \cite{SS}) by less 
than 15\% (using the experimental value of 9.42 mJ/mol K$^2$ \cite{veal}).
\end{references}
\end{document}